\documentstyle[12pt,epsfig]{article}
\def\beq{\begin{equation}}
\def\eeq{\end{equation}}
\def\bea{\begin{eqnarray}}
\def\eea{\end{eqnarray}}
\def\non{\nonumber}

\def\d{\partial}

\begin{document}

% \begin{flushright} hep-th/9701097

% \end{flushright}

% \rightline{TIFR/TH/97-01}

% \rightline{January 1997}

\begin{center}

{\large \bf \sf Phase shift analysis of PT-symmetric nonhermitian
extension of $A_{N-1}$ Calogero model without confining
interaction }

\vspace{1.3cm}

{\sf B. Basu-Mallick$^1$\footnote{e-mail address:
biru@theory.saha.ernet.in}, Tanaya
Bhattacharyya$^1$\footnote{e-mail address:
tanaya@theory.saha.ernet.in} and Bhabani Prasad
Mandal$^2$\footnote{e-mail address: bpm@bose.res.in}}

\bigskip
{\em $^1$Theory Group, Saha Institute of Nuclear Physics, \\ 1/AF
Bidhan Nagar, Kolkata 700 064, India}

\bigskip

{\em $^2$Department of Physics, Maulana Azad College, \\ 8, Rafi
Ahmed Kidwai Road, Kolkata 700013, India}

\bigskip

\noindent {\bf Abstract}

\end{center}

We discuss a many-particle quantum system, which is obtained by
adding some nonhermitian but PT (i.e. combined parity and time
reversal) invariant  interaction to the $A_{N-1}$ rational
Calogero model  without confining potential. This model gives rise
to scattering states with continuous  real spectrum. The
scattering phase shift is determined through the exchange
statistics parameter. We find that, unlike the case of usual
Calogero model, the exclusion and exchange statistics parameter
differ from each other in the presence of PT invariant
interaction.

\newpage

\noindent \section{Introduction}
\renewcommand{\theequation}{1.{\arabic{equation}}}
\setcounter{equation}{0}

\medskip

 The subject of exactly solvable
 many particle quantum mechanical systems with long-range interactions
is closely connected to diverse subjects like fractional
statistics, random matrix theory, level statistics for disordered
systems, Yangian algebra etc. and generated a lot of interest in
past years. The $A_{N-1}$ Calogero model (related to $A_{N-1}$ Lie
algebra) is the simplest example of such a dynamical model,
describing $N$ particles on a line and with Hamiltonian given by
\cite {ca} 
\beq
 H= -  {1\over 2} \sum_{j=1}^N {\d^2 \over \d
x_j^2} + {g \over 2} \sum_{j\neq k} {1 \over (x_j -x_k)^2}
 + {\omega^2\over 2} \sum_{j=1}^N x_j^2 \, ,
\label{1} 
\eeq
 where $g$ is the coupling of long-range interaction and 
 $\omega$ is the coupling of harmonic confining interaction.  
This $A_{N-1}$ Calogero model can be
solved exactly to obtain  the complete set of discrete energy eigenvalues
and corresponding bound state eigenfunctions. 
The complete set of energy eigenvalues can be
written as 
\beq 
E_{n_1,n_2,\cdots n_N} =
\frac{N\omega}{2}[1+(N-1)\nu] +\omega\sum_{j=1}^N n_j \, ,
\label{ee}
\eeq
where $n_j$s are non-negative integer valued quantum numbers
with $n_j\leq n_{j+1}$ and $\nu$ is a real positive parameter related
to the coupling constant of long range interaction as 
\beq 
g=\nu^2
-\nu. 
\label{gg} 
\eeq 
This spectrum in (\ref{ee}) is same as that
for a $N$ number of free bosonic oscillators apart from a constant
shift for all energy levels. This spectrum can be expressed exactly in the
form of the energy eigenvalues of free bosonic oscillators : $
E_{n_1,n_2,\cdots n_N} = \frac{N\omega}{2}
+\omega\sum_{j=1}^N\bar{n}_j $,  where $ \bar{n}_j \equiv n_j +\nu(j-1)
$ are quasi-excitation numbers. These quasi-excitation numbers are
no longer integers and they satisfy a modified selection rule
given by  $ \bar{n}_{j+1} -\bar{n}_j \ge \nu$ . This selection
rule restricts the difference between the quasi-excitation numbers
to be at least $\nu$ apart. As a consequence, the Calogero model
(\ref{1}) provides a microscopic realization for generalized
exclusion statistics (GES) \cite{ha} with $\nu$ representing the
corresponding GES parameter \cite{is,po,bk}.

 The Calogero model without confining potential, which is obtained by
setting $\omega =0 $ in eqn.(\ref{1}), is also studied in Ref.[1]. 
Unlike the earlier case, the spectrum of this model is
continuous and only scattering states occur. Due to such
scattering, particle momenta in a outgoing $N$-particle plane wave
get rearranged (reversely ordered) in terms of the momenta in the
incoming plane wave. The corresponding momentum independent
 scattering phase shift is
given by $ \theta_{sc}= \pi\nu\frac{N(N-1)}{2}$, which is simply
$\nu \pi$ times the total number of two-body exchanges that is
needed for rearranging $N$ particles in the reverse order. Thus it
is natural to identify $\nu$ as the exchange statistics parameter
in this case \cite{po}. It may be noted that this exchange
statistics parameter coincides with the exclusion statistics
parameter as obtained earlier in the presence of confining
potential.

In the last few years, theoretical investigations on different
nonhermitian Hamiltonians have received a major boost because many
such systems, whenever they are invariant under combined parity
and time reversal ($PT$) transformation, lead to either real (when $PT$
symmetry is unbroken)
  or pairs of complex conjugate energy eigenvalues (when $PT$ symmetry is
spontaneously  broken) \cite{pt1,pt2,pt3,ali,wei}. Such property
of energy eigenvalues in nonhermitian $PT$ invariant
 systems can be related to
 the pseudo-hermiticity \cite{ali} or anti-unitary symmetry \cite{wei}
of the corresponding Hamiltonians.
As concrete examples of $PT$ symmetric quantum mechanics,
the Hamiltonians of only one particle in one space dimension have
mostly been considered in the literature so far. However,
nonhermitian but
PT invariant extension of some exactly solvable many particle
quantum mechanical system in one space dimension have also been
considered recently
\cite{bm,tia,bri,Fei}. The PT transformation for such $N$-particle
system can be written as 
\beq 
i\rightarrow-i, \ \
x_j\rightarrow -x_j, \ \ p_j\rightarrow p_j 
\label{pt}
\eeq 
where
$j\in [1,2\cdots N] $ and $x_j \ (p_j\equiv
-i\frac{\partial}{\partial x_j})$ denotes coordinate (momentum)
operator of the $j-$th particle. In particular, an extension 
of $A_{N-1}$ Calogero model
with confining term is proposed by adding a momentum 
dependent long-range interaction 
($\delta\sum_{j\neq k}\frac{1}{(x_j-x_k)}\frac{\partial}{\partial x_{j}}$)
to the Hamiltonian (\ref{1}):
\beq 
H_{ext}= -  {1\over 2} \sum_{j=1}^N {\d^2 \over \d
x_j^2} +  {g \over 2} \sum_{j\neq k} {1 \over (x_j -x_k)^2} +
\delta\sum_{j\neq k}\frac{1}{(x_j-x_k)}\frac{\partial}{\partial x_{j} } 
+ {\omega^2\over 2} \sum_{j=1}^N x_j^2 \, ,
\label{ext1}
\eeq 
where $\delta $ is a real parameter \cite{bk,tia}. 
 It is shown that this
nonhermitian, PT invariant model
can be solved exactly and within certain range of the related
parameters it yields a real spectrum 
\beq 
E_{n_1 n_2\cdots n_N} =
\frac{N\omega}{2}[1+(N-1){\tilde \nu}
+\omega \sum_{j=1}^{N}n_j\, .
\label{e2}
\eeq 
Here ${\tilde \nu} = {\nu}^\prime - \delta$ and 
${\nu}^\prime $ is a real positive parameter which is related to the coupling
constants $g$ and $\delta$ as  
\beq
 g= {\nu'}^2-{\nu}^\prime \, (1+2\delta) \, .
\label{prime}
\eeq
In analogy with the case
of original Calogero model, the energy eigenvalues (\ref {e2}) 
can be rewritten exactly in the form of energy spectrum
for $N$ free oscillators:
$E_{n_1 n_2\cdots n_N} = \frac{N\omega}{2}
+\omega\sum_{j=1}^{N}\bar{n}_j$, 
where $ \bar{n}_j \equiv n_j + {\tilde \nu}(j-1)$ 
are quasi excitation numbers 
  satisfying a modified selection rule given by
$\bar{n}_{j+1}-\bar{n}_j\geq\tilde{\nu} \, $. 
Consequently, 
the extended Calogero model (\ref{ext1}) 
also provides a microscopic realization for GES, 
where ${\tilde \nu}$ represents the exclusion statistics parameter.

In this context, one can naturally ask 
whether the exclusion and exchange statistics parameters are same
or not in the case of above described 
nonhermitian, PT invariant 
extension of the Calogero model. 
This motivates us to find out the exchange statistics parameter  
associated with such
extension of Calogero model in the absense of confining interaction, 
 whose Hamiltonian is obtained by putting $\omega = 0$ in eqn.(\ref {ext1}). 
In Sec.2 of this article, we briefly review the construction of
 scattering eigenstates and phase shift for the
 $A_{N-1}$ Calogero model in absence of confining
potential. By following the same procedure,  
in Sec.3 we find out the scattering eigenstates for 
the extended Calogero model without confining potential
 and also calculate the related phase shift. 
Using this scattering phase shift, 
we determine the exchange statistics parameter 
for the extended Calogero model.
Section 4 is kept for conclusions.

\vspace{1cm}

\noindent \section{ Scattering states of Calogero model without
confining interaction}
\renewcommand{\theequation}{2.{\arabic{equation}}}
\setcounter{equation}{0}

\medskip

In this section  we briefly discuss the scattering state solutions
and corresponding phase shift of  $A_{N-1}$ Calogero model in the
absence of confining interaction. Such a Calogero model is
described by the Hamiltonian  given in (\ref{1}) with $\omega =0$
as 
\beq
 {\cal H }_0
= -\frac{1}{2}\sum_{j=1}^N \frac{\partial^2}{\partial x_j^2} +
\frac{g}{2} \sum_{j\neq k}\frac{1}{(x_j - x_k)^2}\,.
 \label{18}
\eeq
Following \cite{ca}, the eigenvalue problem
for the above Hamiltonian can be solved to obtain scattering 
states within a sector of
configuration space corresponding to a definite ordering of
particles like $x_1 \geq x_2 \geq \cdots \geq x_N$. 
The zero energy ground state
wavefunction of this model is given by 
\beq 
\psi_{gr}=\prod_{j<k}
(x_j-x_k)^\nu  , 
\label {b1} 
\eeq 
where $\nu$ is a real 
positive parameter satisfying the relation (\ref{gg}).

Next, we consider the general
eigenvalue equation associated with the Hamiltonian (\ref {18}):
\bea 
{\cal H}_0\psi = p^2 \psi, 
\label{ham}
\eea 
where $p$ is real and positive.
Solutions of this eigenvalue equation can be written 
in the form
\beq 
\psi= \psi_{gr}\tau( x_1, x_2, \cdots x_N)\,,
\label{b3} 
\eeq 
where $\tau( x_1, x_2, \cdots x_N)$ satisfies the
following differential equation, 
\beq
-\frac{1}{2}\sum_{j=1}^N\frac{\partial^2\tau}{\partial x_j^2} -
\nu \sum_{j \neq k}\frac{1}{(x_j-x_k)}\frac{\partial
\tau}{\partial x_j} = p^2 \tau \, . 
\label{b4} 
\eeq
 Further, to separate the `radial' and `angular' part of the eigenfunction,
one assumes that $ \tau( x_1, x_2, \cdots x_N)= P_{k,q}(x)\chi(r)$,
where the radial variable $r$ is defined as 
\beq
r^2=\frac{1}{N}\sum_{i \neq j}{(x_i-x_j)}^2 
\label{b5} 
\eeq 
and $P_{k,q}(x)$s are translationally invariant, symmetric, k-th order
homogeneous polynomials satisfying the differential equations 
\beq
\sum_{j=1}^N \frac{\partial^2 P_{k,q}(x)}{\partial x_j^2} +
\nu\sum_{j \neq
k}\frac{1}{(x_j-x_k)}\left(\frac{\partial}{\partial x_j} -
\frac{\partial} {\partial x_k}\right) P_{k,q}(x) = 0\, .
\label{b6} 
\eeq 
Note that the index $q$ in $P_{k,q}(x)$ can take
any integral value ranging from $1$ to $g(N,k)$, where $g(N,k)$ is
the
 number of independent polynomials
which satisfy eqn.(\ref{b6}) for a given $N$ and $k$ \cite{ca}.
One should further note at this point that the translational
invariance and homogeneity property of the polynomial
$P_{k,q}(x)$ enforce it to satisfy the following relations, 
\bea
\sum_{j=1}^N \frac{\partial P_{k,q}(x)}{\partial x_j} = 0 \, ,
~~~~~~~~ \sum_{j=1}^N x_j\frac{\partial P_{k,q}(x)}{\partial x_j}
= kP_{k,q}(x) \, . 
\label{prop} 
\eea 
By substituting $P_{k,q}(x)\chi(r)$ in the place of 
$ \tau( x_1, x_2, \cdots x_N)$ 
  in eqn.(\ref{b4}), and using the relations (\ref{b6}), 
(\ref{prop}),  one obtains 
 \beq
-\frac{d^2 \chi(r)}{d r^2} -
\frac{(1+2b)}{r} \, \frac{d \chi(r)}{d r}= p^2 \chi(r)
\, , 
\label{b9} 
\eeq 
where $b=\frac{N-3}{2} + k +
\frac{N(N-1)\nu}{2}$. The above equation admits a
solution of the form $\chi(r)= r^{-b}J_b(pr)$, where $J_b(pr)$
denotes the Bessel function. Hence the scattering state
 eigenfunctions of ${\cal H}_0$ with eigenvalue $p^2$
  are finally obtained as
\beq 
\psi =\prod_{j<k}{(x_j-x_k)}^{\nu}r^{-b} J_b(pr) P_{k,q}(x)\, .
\label{sol1} 
\eeq

Next we want to discuss the scattering phase shift for
the above model. For this purpose, one has to construct 
a more general eigenfunction which 
  in the asymptopic limit (i.e., $r \rightarrow \infty$ limit)
  can be expressed in terms of an incoming free
particle wavefunction
($\psi_{+}$) and an outgoing free particle wavefunction
 ($\psi_{-}$),  where the incoming
wavefunction will be of the form 
\beq 
\psi_{+}= \exp[\,i\sum_{j=1}^N p_j
x_j\,] \, , 
\label{b12} 
\eeq 
with $p_j \leq p_{j+1}$, $p^2
=\sum_{j=1}^N p_j^2 $ and $\sum_{j=1}^N p_j = 0. $ 
This can be achieved by taking appropriate linear superposition of 
all degenerate eigenfunctions (with eigenvalue $p^2$) of
the form (\ref{sol1}):
\beq 
\psi_{gen} =
\prod_{j<k}{(x_j-x_k)}^{\nu}\sum_{k=0}^{\infty}\sum_{q=1}^{g(N,k)}
C_{kq} r^{-b} J_b(pr) P_{k,q}(x)\, , 
\label{b10} 
\eeq 
where
$C_{kq}$s are expansion coefficients depending on particle
momenta. We assume that each of the momenta $p_i$ is a product of
a radial part $p$ and an angular part $\alpha_i$, i.e, $p_i =
p\alpha_i$. Now by matching the dimensions of the right hand sides of 
equations (\ref {b12}) and (\ref {b10}), it can be shown that
$C_{kq} = p^{3-N\over 2}
{\tilde C}_{kq} (\alpha_i)$, where 
${\tilde C}_{kq} (\alpha_i)$ depends only on the angular parts 
of the momenta.
By using this relation for $C_{kq}$ and also the
asymptotic properties of Bessel function at
 $r \rightarrow \infty $ limit,
$$J_b(pr) \rightarrow \frac{1}{\sqrt{2\pi
pr}}\{e^{-i(n+\frac{1}{2})\frac{\pi}{2} + ipr} +
e^{i(n+\frac{1}{2})\frac{\pi}{2} - ipr}\}\, , $$ one can write
down the asymptotic from of 
  $\psi_{gen}$ (\ref {b10}) as 
\beq 
\psi_{gen} \sim \psi_{+}
+ \psi_{-} \, , 
\label{b11}
 \eeq 
where \beq \psi_{\pm} = {(2\pi
r)}^{-\frac{1}{2}}p^{(n-\frac{1}{2})}\prod_{j<k}{(x_j-x_k)}^ {\nu}
r^{-A}\sum _{k=0}^{\infty}\sum_{q=1}^{g(N,k)}{\tilde
C}_{kq}(\alpha_i) r^{-k}P_{k,q}(x) e^{\pm i(b +
\frac{1}{2})\frac{\pi}{2}\mp ipr}, 
\label{b101} 
\eeq
 with $A
= b-k =\frac{N-3}{2} +\frac{N(N-1)\nu}{2} $ and $n=
\frac{3-N}{2}$. Now, to get the expression for the phase shift due
to scattering,  the outgoing wavefunction $\psi_-$  has to be
expressed in terms of the incoming wavefunction $\psi_+$. For this 
purpose let us consider a special permutation of the particle coordinates
$T$, defined as $$T x_i = x_{N-i+1}~,~~~~~~i=1,2, \cdots,N$$ such
that the set $\{-Tx\}$ belongs to the same sector as that of the
set $\{x\}$, with the particle ordering $x_1 \geq x_2 \geq \cdots
\geq x_N$. It is also to be noted here that the symmetry and the
homogeneity of $P_{k,q}(x)$ enables us to write the relation,
$$P_{k,q}(-Tx) = e^{-ik\pi}P_{k,q}(x).$$ Hence, by taking 
advantage of the above
facts and finally using eqn.(\ref {b12}), 
it can be shown that the outgoing wavefunction $\psi_-$
can be written as 
\bea 
\psi_{-} &=& e^{-i\pi(A+n)} \psi_+
\Big(x\rightarrow -Tx,~ p\rightarrow -p \Big) \non \\ 
&=&e^{-i\pi\nu \frac{N(N-1)}{2}}\exp[\,i\sum_{j=1}^N x_{j}
\,p_{N+1-j}\,] \, . 
\label{b13} 
\eea 
Comparing (\ref {b12}) with
(\ref {b13}), one finds that
 the momenta of the incoming plane wave gets rearranged (reversely ordered)
in the scattering process. Moreover, the outgoing plane wave acquires 
a momentum independent phase shift 
given by $\pi\nu
 \frac{N(N-1)}{2}$.  Thus $\nu$ can be identified with the exchange
statistics parameter associated with this Calogero model.

\noindent\section{ Scattering phase shift of extended Calogero
Model without confining interaction}

Here we aim to study the $PT$ symmetric nonhermitian extension of
the Calogero model without confining interaction. We extend the
Hamiltonian (\ref{18}) by adding an extra term $ \delta
\sum_{j\neq k}\frac{1}{(x_j-x_k)}\frac{\partial}{\partial x_j},$
which is nonhermitian but symmetric under combined $PT$
transformation. Hence the extended Calogero model which we will be
considering here is described by the Hamiltonian 
\beq
{\cal H}_{ext} = -\frac{1}{2}\sum_{j=1}^N\frac{\partial^2}{\partial
x^2_j}+\frac{g}{2}\sum_{j\neq}\frac{1}{(x_j-x_k)^2}+\delta
\sum_{j\neq k}\frac{1}{(x_j-x_k)}\frac{\partial}{\partial x_j}.
\label{hh} 
\eeq 
Since $H_{ext}$ (\ref{ext1}) reduces to the above Hamiltonian at 
$\omega = 0$ limit, the exchange statistics parameter of extended
Calogero model can be determined from the phase shift analysis of the
scattering states associated with the Hamiltonian (\ref{hh}).
Now we follow the same procedure as in the
previous section to study the scattering states of this extended
model.

We start by observing that the 
zero energy ground state wave function of 
 the Hamiltonian (\ref{hh}) is quite similar in form 
 with the  ground state wave function (\ref{b1}) of the original Calogero 
model:
\beq 
\psi_{gr}=\prod_{j<k}
(x_j-x_k)^{\nu'}  \, , 
\label {gr} 
\eeq 
 where the modified exponent 
 ${\nu}^\prime$ is related to the coupling
constants $g$ and $\delta$ through the relation (\ref {prime}).
For the purpose of obtaining nonsingular ground state eigenfunction
at the limit $x_i \rightarrow x_j$, $\nu^\prime$ should be a non-negative
exponent. Due to eqn.(\ref {prime}), this condition 
restricts the ranges of coupling constants $g$ and $\delta$ as 
(i) $\delta \geq - \frac{1}{2}, ~0>g\geq-(\delta +\frac{1}{2} )^2,$
and (ii) $g \geq 0$ with arbitrary value of $\delta$. 
Next, we consider the general
eigenvalue equation associated the Hamiltonian (\ref {hh}) given by
\bea 
{\cal H}_{ext} \, \psi \, = \,  p^2 \, \psi ,
\label{gen}
\eea 
where $p$ is a real positive parameter.
It is easy to see that the solutions of this eigenvalue equation
 can be written in the form $ \psi =\psi_{gr}
\tau^\prime (x_1, x_2 \cdots x_N)$, where 
 $\psi_{gr}$ represents the modified
  ground state eigenfunction (\ref {gr}) and 
$\tau^\prime(x_1,x_2 \cdots x_N)$
 satisfies a differential equation like 
\beq
-\frac{1}{2}\sum_{j=1}^N\frac{\partial^2\tau^\prime}{\partial x_j^2} -
({\nu}^\prime - \delta) \sum_{j \neq
k}\frac{1}{(x_j-x_k)}\frac{\partial \tau^\prime}{\partial x_j} = 
p^2 \tau^\prime \, . 
\label{tau} 
\eeq
%The above
%equation is similar to eqn.(\ref{b4}), except that $\nu$ is now
%replaced by $\nu - \delta$.
Next we assume that 
$\tau^\prime(x_1 , x_2 \cdots x_N) $ can be factorised as 
\beq
\tau^\prime(x_1 , x_2 \cdots x_N) = P_{k,q}^\prime (x)
\chi^\prime(r), 
\label{taup}
\eeq
where $r$ is the radial variable  defined in (\ref{b5}) and 
$P_{k,q}^\prime(x)$s are translationally invariant, symmetric, k-th
order homogeneous polynomials satisfying the differential
equations 
\beq 
\sum_{j=1}^N \frac{\partial^2 P_{k,q}^\prime(x)}{\partial
x_j^2} + ({\nu}^\prime-\delta)\sum_{j \neq
k}\frac{1}{(x_j-x_k)}\left(\frac{\partial}{\partial x_j} -
\frac{\partial} {\partial x_k}\right) P_{k,q}^\prime(x) = 0\, .
\label{poly} 
\eeq 
Note that the form of eqn.(\ref{poly}) is same as
eqn.(\ref{b6}) apart from the fact that here
 $\nu$ is replaced by $\nu^\prime
- \delta $. Hence it is clear that $P_{k,q}^\prime (x)$ can be obtained from
any given expression of $P_{k,q}(x)$ 
by simply substituting the parameter $\nu$ with $\nu^\prime - \delta $.
So the index $q$ in $P_{k,q}^\prime (x)$ can also take values ranging from 1
to $g(N,k)$.
Substituting the factorised form (\ref {taup}) of $\tau(x_1 ,x_2
\cdots x_N)$ in the differential eqn.(\ref{tau}) and making use of
the properties of $P_{k,q}^\prime(x)$ we obtain the equation satisfied by
the `radial' part of the wavefunction as 
\beq
-\frac{\partial^2\chi^\prime(r)}{\partial r^2}
-\frac{1+2b^\prime}{r}\frac{\partial\chi^\prime(r)}{\partial r} =
p^2\chi^\prime(r) 
\label{bb} 
\eeq 
with $ b^\prime = \frac{N-3}{2} +k +
({\nu}^\prime -\delta)\frac{N(N-1)}{2}$.
 %\equiv A^\prime +k$, where
%$A^\prime =A -\delta\frac{N(N-1)}{2}$.
The solution of eqn.(\ref{bb}) can be expressed through the Bessel 
function: $\chi^\prime
(r) = r^{-b^\prime}J_{b^\prime}(pr)$. Hence the scattering state
eigenfinctions of ${\cal H}_{ext}$ (\ref {hh})
 with real positive eigenvalue $p^2$ are obtained as
\beq 
\psi =\prod_{j<k}{(x_j-x_k)}^{\nu^\prime}r^{-b^\prime} J_{b^\prime}(pr) 
P_{k,q}^\prime(x)\, .
\label{sol2} 
\eeq 

Next, we carry out the calculation of phase shift in the
same manner as in the previous section. First, we aim to construct
a more general eigenfunction of ${\cal H}_{ext}$ such that 
in the asymptotic limit it can be
expressed in terms of an incoming wave ($\psi_+$) of the form (\ref{b12}) 
and an outgoing wave ($\psi_-$). For this purpose 
one has to take appropriate linear superposition of all degenerate
eigenfunctions (with eigenvalue $p^2$) of the form (\ref{sol2}):
\beq
{\psi}_{gen}=
\prod_{j<k}{(x_j-x_k)}^{{\nu}^\prime}\sum_{k=0}^{\infty}\sum_{q=1}^{g(N,k)}
C^{\prime}_{kq} r^{-b^\prime} J_{b^\prime}(pr) P_{k,q}^\prime(x)\, ,
\label{b20} 
\eeq 
where $C^\prime_{kq}$s are expansion coefficients
which are functions of particle monenta. Once again by doing 
dimensional analysis, we obtain 
$$
C^\prime_{kq} \, = \,  p^{\frac{(3-N)}{2}+
\frac{N(N-1)\delta}{2}} \, {\tilde C}_{kq}^\prime(\alpha_i) \, ,
$$
where ${\tilde C}_{kq}^\prime(\alpha_i)$ depends only on the angular parts
of the momenta.
% $C^{\prime}_{kq}(p,\alpha_i) =
%p^{n^\prime}{\tilde{C}}^{\prime}_{kq}(\alpha_i)$ where $n^\prime =
%\frac{3-N}{2} + \frac{N(N-1)\delta}{2} \equiv n+
%\frac{N(N-1)\delta}{2}$.
By using this explicit expression for $C^\prime_{kq}$ and the 
asymptotic properties of Bessel function at $r
\rightarrow \infty$, we obtain the asymptotic form of 
${\psi}_{gen}$ (\ref{b20}) as 
$ {\psi}_{gen} \sim {\psi}_{+} + {\psi}_{-} ,$ 
where $$ {\psi}_{\pm} = {(2\pi
r)}^{-\frac{1}{2}}p^{({n}^\prime-\frac{1}{2})}
\prod_{j<k}{(x_j-x_k)}^ {{\nu}^\prime}r^{-A^\prime}\sum
_{k=0}^{\infty}\sum_{q=1}^{g(N,k)}{{\tilde
C}^\prime}_{kq}(\alpha_i)
 r^{-k}P_{k,q}^\prime(x)
e^{\pm i(b^\prime + \frac{1}{2})\frac{\pi}{2}\mp ipr}. $$ In the
above expression $A^\prime = b^\prime -k = \frac{N-3}{2}
+(\nu^\prime-\delta)\frac{N(N-1)}{2} $ and $n^\prime =
\frac{3-N}{2} + \frac{N(N-1)\delta}{2}$.
Following the same procedure as in the
case of $A_{N-1} $ Calogero model without confining term, we find
that the outgoing wavefunction $(\psi_-) $ can be written as 
\bea
\psi_{-} &=& e^{-i\pi(A^\prime+n^\prime)} \psi_+ \Big(x\rightarrow
-Tx,~ p\rightarrow -p \Big) \non \\
&=&e^{-i\pi{\nu}^\prime\frac{N(N-1)}{2}}\exp[\,i\sum_{j=1}^N
x_{j} \,p_{N+1-j}\,] \, . 
\label{b23} 
\eea 
Comparing eqns.(\ref {b23}) and (\ref {b12}),  we observe that
 the momenta of the incoming plane wave gets
rearranged (reversely ordered) in the scattering process. 
Furthermore, the outgoing plane wave now acquires a momentum independent
phase shift given by $\pi\nu^\prime \frac{N(N-1)}{2}$, which is 
$ \pi \nu^\prime $  times the number of two body exchanges needed
for rearranging $N$-particles in the reverse order.
Thus $\nu^\prime$ can be identified with the exchange
statistics parameter associated with this extended Calogero model.

\noindent\section
{Conclusion}

Here we study a nonhermitian but $PT$ invariant extension
of the $A_{N-1}$ Calogero model without confining interaction.
Unlike the case of $A_{N-1} $ Calogero model with
confining term, here we find only scattering states with
continuous, real eigenvalues. We also calculate the scattering
phase shift explicitly for this model. The
exchange statistics parameter for this model is determined through
the scattering phase shift of plane waves. Such 
exchange statistics parameter is given by $\nu^\prime$, which is 
related to the coupling constants through eqn.(\ref{prime}).
 However, it is found earlier that the exclusion statistics parameter
 associated with PT invariant extension of Calogero model in the presence of
confining interaction is given by 
 ${\tilde \nu} \equiv {\nu}^\prime - \delta$.
 Thus we surprisingly find that,
in contrary to the case of original Calogero model,
 the exclusion and exchange statistics parameters
  differ from each other in the
presence of $PT$ invariant interaction. In particular,
 within a range 
of coupling constant given by $\delta >0, ~0>g> -\delta (1+ \delta)$, 
the value of exclusion statistics parameter ${\tilde \nu}$
becomes negative. As a result, the ground state energy does not have any
lower bound in the thermodynamic limit (i.e., $N\rightarrow \infty$ 
limit). On the other hand, the exchange statistics parameter 
$\nu^\prime$ is positive
and well defined even within the above mentioned range of the coupling
constants.

\end{document}